\documentclass[twocolumn,showpacs,preprintnumbers,aps,prl,amsmath,amssymb,floatfix]{revtex4}

\usepackage{graphicx}
\usepackage{dcolumn}
\usepackage{bm}

\begin{document}

\title{Photoassociative Production and Trapping of Ultracold KRb Molecules}

\author{D. Wang$^1$, J. Qi$^1$, M. F. Stone$^1$, O. Nikolayeva$^2$, H. Wang$^3$}
\author{B. Hattaway$^1$, S. D. Gensemer$^4$}
\author{P. L. Gould$^1$}
\author{E. E. Eyler$^1$}
\author{W. C. Stwalley$^1$}
\affiliation{$^1$Physics Department, University of Connecticut,
 Storrs, CT 06269}
\affiliation{$^2$Physics Department, University of Latvia, 19 Rainis
Boulevard, Riga, Latvia 15986} \affiliation{$^3$The Aerospace
Corporation, M2-253, 2350 East El Segundo Boulevard, El Segundo, CA
90245-4691}
 \affiliation{$^4$Van der Waals-Zeeman Institut, Universiteit van Amsterdam,
 Valckenierstraat 65, 1018 XE Amsterdam, The Netherlands }
\date{\today}

\begin{abstract}
We have produced ultracold heteronuclear KRb molecules by the
process of photoassociation in a two-species magneto-optical trap.
Following decay of the photoassociated KRb*, the molecules are
detected using two-photon ionization and time-of-flight mass
spectroscopy of KRb$^+$. A portion of the metastable triplet
molecules thus formed are magnetically trapped. Photoassociative
spectra down to 91 cm$^{-1}$ below the K(4$s$) + Rb (5$p_{1/2}$)
asymptote have been obtained. We have made assignments to all eight
of the attractive Hund's case (c) KRb* potential curves in this
spectral region.
\end{abstract}

\pacs{33.80.Ps, 32.80.Pj, 33.20.-t,34.50.Gb  \hspace{20 mm}Accepted
by \textit{Physical Review Letters}}

\maketitle

Following in the footsteps of ultracold atoms, the study of
ultracold molecules has seen significant progress in recent years
\cite{Bahns00}. Homonuclear molecules have been produced at
ultracold temperatures, here defined as $T < 1 $ mK, by
photoassociating ultracold atoms \cite{Stwalley99,Masnou01}. In a
few cases, the resulting molecules have been subsequently trapped
\cite{Takekoshi98,Vanhaecke02}. Feshbach resonances in even colder
atomic samples have yielded very weakly bound homonuclear molecules
\cite{Feshbach}, in some cases under quantum degenerate conditions.
Heteronuclear polar molecules have attracted particular interest
because their permanent dipole moments can be controlled with
external fields. At ultracold temperatures, the resulting
long-range, anisotropic dipole-dipole interactions are expected to
lead to new phenomena, such as anisotropic collisions \cite{Bohn01}
and novel quantum degenerate behavior \cite{Santos00, Yi00,Goral02}.
Heteronuclear photoassociative spectroscopy \cite{Wang98} is also of
interest, and applications of ultracold polar molecules to quantum
computing \cite{DeMille02} and fundamental symmetries
\cite{Kozlov95} are also being considered.

\begin{figure}
\centering
\includegraphics[width=0.95\linewidth]{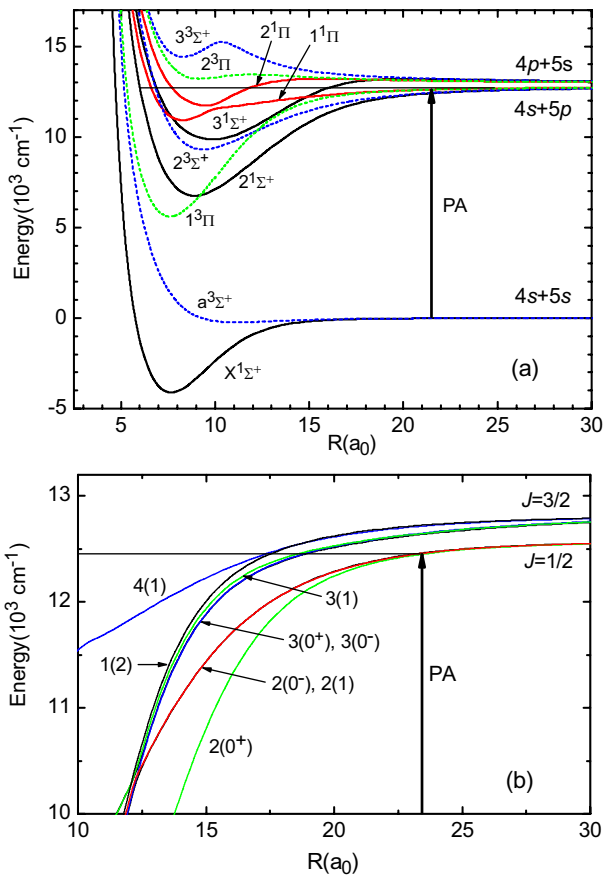}
 \caption{\protect\label{fig1} The
photoassociation process (PA) and possible singlet (X$^1\Sigma^+$)
and triplet (a$^3\Sigma^+$) molecule formation processes in KRb
based on the \textit{ab initio} potential energy curves of Ref.
\cite{Rousseau00}.  In (b), the long-range states converging to 4$s$
+ 5$p_J$ are shown; the state designations consist of an ordinal
numbering followed by the Hund's case (c) symmetry in parentheses,
expressed as $\Omega^{\pm}$ . The long-range states converging to
$4p_J + 5s_{1/2}$ (not shown) are all repulsive.}
\end{figure}

In this Letter, we report on the production, detection, magnetic
trapping, and photoassociation spectroscopy of KRb molecules.  This
work builds on a variety of other recent advances, particularly the
recent production and spectroscopic study of  ultracold polar RbCs
molecules by Kerman and coworkers \cite{Kerman04a,Kerman04b}, using
photoassociation (PA) in overlapping Rb and Cs magneto-optical traps
(MOTs). Similar PA spectroscopy has been performed with the
heteronuclear (but nonpolar) $^6$Li$^7$Li molecule
\cite{Schloder01}. In other related work, observations of ultracold
NaCs$^+$ \cite{Shaffer99}, RbCs \cite{Wang03}, and KRb
\cite{Mancini04} have been reported in two-species MOTs, although
the formation mechanisms have not been identified. Heteronuclear
Fesh\-bach resonances have been recently reported in Li+Na
\cite{Stan04} and K+Rb \cite{Inouye04}, although the resulting
molecules were not directly observed. At temperatures above 10 mK,
polar molecules have been successfully cooled and trapped.
 CaH has been buffer-gas cooled and magnetically trapped
\cite{Weinstein98}, while ND$_3$ has been electrostatically slowed
and loaded into electric traps \cite{Bethlem00,Crompvoets01}.

Photoassociation, as shown in Figure 1, involves the free-bound
absorption of a photon by a colliding pair of atoms
\cite{Stwalley99,Masnou01}.  It is a particularly powerful
spectroscopic technique at $\mu$K temperatures, where only a few
partial waves contribute, as the centrifugal barriers of collision
are significantly greater than the thermal kinetic energy for quite
low collisional angular momenta ($\ell\lesssim{3}$).  Thus the bound
states formed are also limited to low rotational angular momenta
($J\lesssim4$ ).

In our experiments, electronically excited KRb* molecules are
produced by PA in overlapping K and Rb MOTs, followed by radiative
stabilization via decay to the ground $X ^1\Sigma^+$ state or the
metastable $a ^3\Sigma^+$ state.  Dark-spot MOTs are used to achieve
high atomic densities, and ionization detection is employed for
improved sensitivity. We are able to observe trapping of ultracold
heteronuclear molecules for the first time, by confining triplet KRb
molecules in the inhomogeneous magnetic field of the MOT. We also
present high-resolution photoassociative spectra down to 91
cm$^{-1}$ below the K(4\textit{s}) + Rb (5$p_{1/2}$) asymptote.  In
these spectra we have identified transitions to all eight of the
attractive long-range potentials converging to 4$s$ + 5$p_{1/2}$ and
4$s$ + 5$p_{3/2}$.

 Our $^{39}$K/$^{85}$Rb dual species MOT \cite{Stone} uses the same
apparatus previously used for trapping $^{39}$K \cite{Wang96} with
the addition of trap and repump beams for $^{85}$Rb.  Frequently the
K and Rb diode laser beams can share optics, because of the
closeness of the resonance energies (see Fig. 1).  Indeed it is this
near degeneracy which makes the KRb $C_6$ values especially large
\cite{Wang98}. Normally we operate each of the atom traps as a
``dark-spot" MOT \cite{Ketterle}, with typical densities of
10$^{11}$/cm$^3$ for Rb and 3 $\times$ 10$^{10}$/cm$^3$ for K. The
two MOTs are carefully aligned to maximize overlap of the two atomic
clouds.

To achieve sensitive and species-selective detection we employ
two-photon ionization at 602.5 nm followed by time-of-flight mass
spectroscopy, an example of which is shown in the inset of Fig. 2.
The ionizing laser pulse (1.5 mJ, $\sim$10 ns) is focused to provide
a typical peak intensity of $7 \times 10^6$ W/cm$^2$. These pulses
are generated using a doubled Nd:YAG laser at 532nm to pump a pulsed
dye laser at 10Hz. Ion signals can also be obtained using a 532 nm
laser directly. A Channeltron is used to detect atomic and molecular
ions. We obtain PA spectra by tuning a cw tunable PA laser (Coherent
829-29, typically $>$ 400 mW) while detecting the KRb$^+$
time-of-flight peak.

\begin{figure}
\centering \vskip 0 mm
\includegraphics[width=0.95\linewidth]{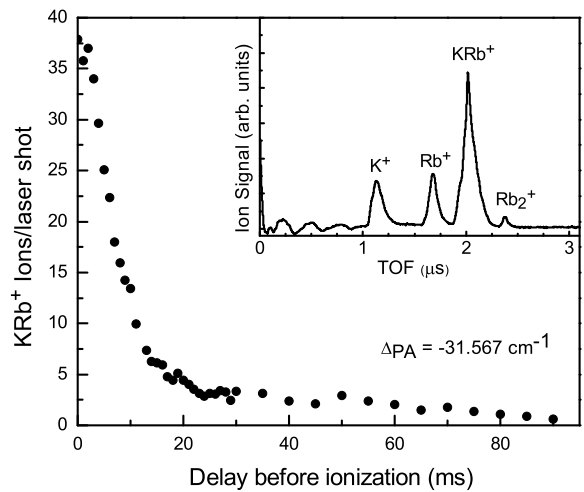}
\caption{\protect\label{fig2} Main panel: Time evolution of the
$^{39}$K$^{85}$Rb$^+$ signal as a function of the delay of the
ionization pulse relative to the turnoff of the PA laser, with the
PA laser set to a 3(0$^-$) state resonance as indicated in Fig.
\ref{fig3}(a). PA laser detuning $\Delta_{PA}$ is measured relative
to K(4$s$, $F$=1) + Rb(5$p_{1/2}$, $F$=2).  Inset: Typical
time-of-flight mass spectrum, with the PA laser tuned to the same
resonance.}
\end{figure}

We obtain typical signals of 10-60 KRb$^+$ ions per laser shot when
the PA laser is tuned to a strong KRb resonance.  Surprisingly,
these signals are comparable to the Rb$_2^+$ ion signals observed
when photoassociating on a strong Rb$_2$ line. The number of
colliding atom pairs available at the Condon radius, where the PA
laser matches the ground-excited free-bound transition, strongly
favors Rb$_2$. Its resonant-dipole 1/$R^3$ potentials permit
photoassociation at much longer range (e.g., 100 $a_0$) compared to
KRb where the van der Waals 1/$R^6$ potential results in
shorter-range excitation (e.g., 30 $a_0$). However, the
Franck-Condon factors for decay to the 1/$R^6$ ground-state
potentials strongly favor KRb. Therefore we expect to produce less
KRb* in the PA step, but with more efficient decay to a bound state.
This is reflected in the barely detectable KRb trap loss signal we
observe in contrast to losses of 10-50\% for Rb$_2$. Also, the
relaxed heteronuclear selection rules allow more PA pathways in KRb
compared to Rb$_2$ \cite{Wang98}. The signal sizes may also be
influenced by different detection efficiencies for the KRb and
Rb$_2$.  Finally, the signal sizes may be enhanced by magnetic
trapping, discussed below. Ignoring this trapping and assuming 100\%
ionization efficiency for KRb and 50\% ion detection efficiency, our
maximum signal size would correspond to a production rate of about
$4 \times 10^4$ KRb molecules/s.

\begin{figure}
\centering \vskip 0 mm
 \includegraphics[width=0.95\linewidth]{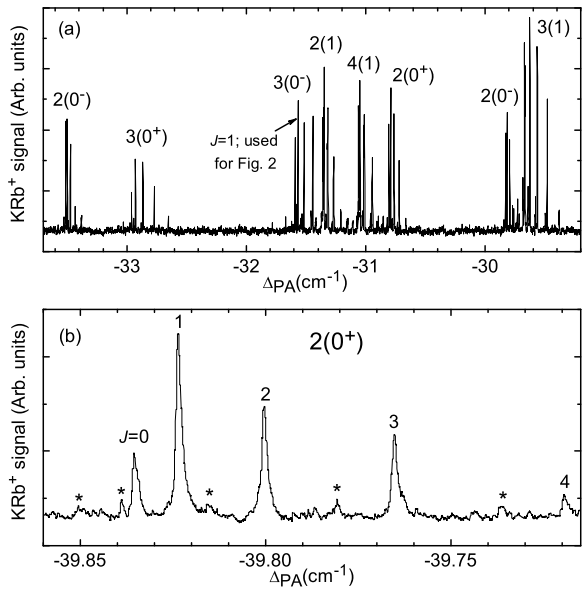}
 \caption{\protect\label{fig3} (a) A PA spectrum for $^{39}$K$^{85}$Rb showing
eight vibrational bands as our PA laser is scanned as indicated
below the K(4s) + Rb(5$p_{1/2}$) asymptote. The tentative state
designation is given for each band using the nomenclature of Fig
1(b). The rotational spacings vary widely, with smaller spacings
corresponding to longer-range states. This spectrum includes
transitions to seven of the eight states correlating with the $J =
1/2$ and $J = 3/2$ asymptotes. All these states can be reached by
dipole-allowed transitions from colliding ground state atoms
($\Omega=0^{{\pm}}, 1$). (b) A high-resolution PA spectrum showing
the rotational structure as the PA laser detuning is scanned through
a 2(0$^+$) state resonance below the K(4s)+ Rb(5$p_{1/2}$)
asymptote. The asterisks(*) indicate ``hyperfine ghosts'' due to a
small population in the ``bright'' $F$=2 state of K(4$s$) in our
dark-spot MOT.}
\end{figure}

    We observe trapping of triplet KRb molecules in
the magnetic field of the MOT coils \cite{Vanhaecke02}.  These
molecules must be in the triplet $a$ state since the singlet
molecules have zero magnetic moment.  We can investigate the
trapping by turning off the PA laser while leaving the MOT lasers
and magnetic field on, then firing the detection laser pulse after a
variable delay time. Figure 2 shows a decay curve observed as a
function of this delay, with the laser tuned to one of the stronger
KRb resonances (the 3(0$^-$) transition marked with an arrow in Fig.
3(a)). The background of 1.7 ions/shot due to KRb formed in the
absence of the PA laser \cite{Mancini04} has been subtracted from
this decay curve. A significant fraction of the molecules survive
long past the ballistic decay time, which is about 3 ms for our 1 mm
diameter detection beam, assuming a temperature of $\sim$300 $\mu$K.
The trapped molecules appear to decay non-exponentially, with a
small signal still observable after 150 ms. Because the atoms in the
MOT are unpolarized, we expect the molecules to have a wide
distribution of projection quantum numbers (m$_J$ or m$_F$,
depending on the coupling scheme), giving rise to a large spread in
the effective trapping potential. For a molecule with the largest
possible magnetic dipole moment projection of two Bohr magnetons
(2$\mu_B = q_e\hbar/m_e$), the horizontal trapping potential reaches
a depth of 500 $\mu$K with a radius of 2.5 mm. The vertical
potential is slightly distorted by gravity, but the magnetic
trapping force is 2.5 times larger than the gravitational force.
Molecules with smaller magnetic dipole moment projections will
experience weaker trapping, and can be expected to be confined in a
larger trap volume, with a higher escape rate and lower detection
efficiency. At least 2/3 of the total molecules formed will be
untrapped or anti-trapped.

\begin{figure}
\centering \vskip 0 mm
 \includegraphics[width=\linewidth]{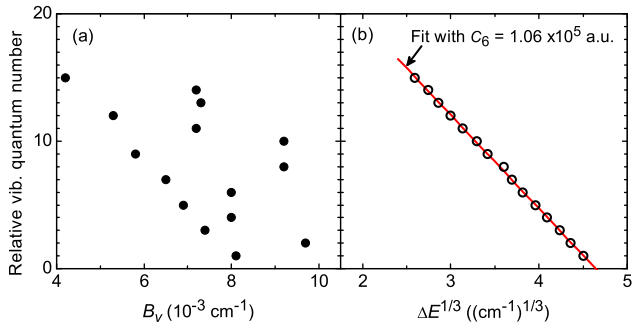}
 \caption{\protect\label{fig4} (a) The rotational constants $B_v$ in cm$^{-1}$ and (b) the
detuning (binding energy) to the 1/3 power for levels of the
2(0$^+$) states converging to the K(4s)+Rb(5$p_{1/2}$) asymptote
versus relative vibrational quantum number (absolute values
unknown). For a pure R$^{-6}$ long-range potential, these plots
should be linear in the absence of perturbations
\cite{LeRoy73,Stwalley78}.}
\end{figure}

    Examples of our heteronuclear PA spectra are given in Figure 3.
The observed linewidths vary from about 35-500 MHz, primarily due to
unresolved internal structure. The excellent signal-to-noise ratio
makes the spectra readily assignable.  At these low $J$ values,
rotational series such as the one shown in Fig. 3(b) are fit
precisely by a linear plot of energy versus ($J(J+1) - \Omega^2$),
yielding a rotational constant $B_v$ without observable centrifugal
distortion. The four 0$^{\pm}$ states (two of each), the three
$\Omega = 1$ states, and the sole $\Omega = 2$ state are readily
assigned in this way. Our $\pm$ assignments are preliminary because
of extensive perturbations \cite{Wang04}. These effects are evident
in Fig. 4, which plots the vibrational energies and $B_v$ values for
one of the 0$^+$ states. We make nominal assignments by associating
the state having the smaller C$_6$ coefficient with the shorter
outer turning point and therefore the larger rotational constant.
Thus the series in Figure 4 is assigned to the 0$^+$ state at the
K(4$s$) + Rb(5$p_{1/2}$) asymptote, i.e. the 2(0$^+$) state of
Figure 1(b). Figure 3(b) in particular corresponds to the ninth
observed vibrational level in the Figure 4 series. There are eight
vibrational levels in Figure 3a, assigned as 0 or 1 states depending
on their lowest rotational quantum number ($J=0$ or 1). One of the 0
levels corresponds to the 11th level in the 2(0$^+$) series in Fig.
\ref{fig4}. Two other 0 levels must correspond to 2(0$^-$), with a
vibrational spacing comparable to 2(0$^+$) at this detuning. The
final two 0 levels, with larger vibrational spacings, are assigned
to the 3(0$^-$) and 3(0$^+$) states, respectively, in order of
increasing $B_v$.

 In a similar way, among the three $\Omega = 1$ levels, one (with $B_v =$ 8.49
$\times$ 10$^{-3}$ cm$^{-1}$) fits well in a long vibrational
series, similar to the 2(0$^+$) series in Fig. 4, and is assigned to
the 2(1) state. The other two levels cannot be in the 2(1) series
and are assigned to the 3(1) and the 4(1) states based on their
relative rotational constants (10.39 and 10.93 $\times$ 10$^{-3}$
cm$^{-1}$, respectively). Since four vibrational levels of the 1(2)
state have also been observed at other detunings, we have been able
to observe all eight of the Hund's case (c) electronic states
correlating to the K(4$s$) + Rb(5$p_{1/2, \;3/2}$) asymptotes, as
shown in Fig 1(b). Further discussion of these assignments will be
published elsewhere \cite{Wang04}.

    Due to our ion extraction configuration, the molecules are formed in a dc field of
$\sim$160V/cm.  We observe no Stark effect at our spectral
resolution of $\sim$0.001 cm$^{-1}$, consistent with the small
dipole moment ($<$0.04 $ea_0$) predicted for KRb*
\cite{Rousseau00,Park03,Kotochigova04,Zemke04}. This differs from
the RbCs case of Ref. \cite{Kerman04a}, where both the field (390
V/cm) and the dipole moment ($\sim$0.5 $ea_0$) are larger.  We note
that the dipole moments of the $X^1\Sigma^+$ and $a^3\Sigma^+$
states of KRb at their equilibrium separations are -0.30(2) and
-0.02(1) $ea_0$, respectively \cite{Kotochigova03}.

    Assuming our assignments are correct for the 2(0$^+$), 2(0$^-$), and 2(1)
states, one can estimate the corresponding effective C$_6$ values
\cite{LeRoy73,Stwalley78} as $(1.06 \pm 0.06) \times 10^5$, $(1.01
\pm 0.04) \times 10^5$ and $(1.01 \pm 0.05) \times 10^5$ atomic
units (a.u.), respectively. Note that $C_6$ is a weak function of
$R$ between very long-range (spin-orbit $\gg$ dispersion) and
intermediate range (dispersion $\gg$ spin-orbit)---the calculations
of \cite{Bussery87}, for example, vary from $1.0 \times 10^5$ a.u.
at very long range to $3.1 \times 10^5$ a.u. at intermediate range
for the 2(0$^+$) state. Model calculations to better define the
dispersion interaction are underway.

    It is important to consider the effects of symmetry on molecule
formation.  Some relevant transition dipole moment functions are
available \cite{Kotochigova03}.  For levels near dissociation, all
of the excited states we produce decay with at least 78\%
probability to the $a ^3\Sigma^+$ state. In fact, the 2(0$^-$),
3(0$^-$), and 1(2) states radiate 100\% to this triplet state. Thus,
since we observe KRb$^+$ from 0$^-$ and 2 states, we know our ion
detection is sensitive to triplet molecules. However, since even the
0$^+$ and 1 states radiate primarily to triplet states at long
range, we do not yet know if our ionization detection is sensitive
to singlet molecules. We hope that in future experiments where both
singlet and triplet molecules are produced but only triplets are
trapped, we can determine the relative
sensitivity for singlet and triplet molecule detection.

In summary, we have significantly advanced the study of ultracold
polar molecules, trapping them for the first time.  KRb molecules
are produced by high-resolution photoassociation and detected by
ionization.  We have extensively explored the PA spectrum,
identifying all eight of the electronic states expected near the
K(4$s$) + Rb(5$p$) limits.

We gratefully acknowledge support from the National Science
Foundation and the University of Connecticut Research Foundation and
helpful discussions with Warren Zemke, Tom Bergeman, David DeMille
and Jamie Kerman, and laboratory assistance from Andrew Scott and Ye
Huang.

\end{document}